\documentclass{article}
\usepackage{amsmath} 
\usepackage{etoolbox}
\patchcmd{\thebibliography}{\section*{\refname}}{}{}{}

\usepackage{amsfonts}
\usepackage[symbol]{footmisc}
\usepackage{cleveref}
\makeatletter
\newcommand\footnoteref[1]{\protected@xdef\@thefnmark{\ref{#1}}\@footnotemark}
\makeatother

\date{Dated: \today}
\title{RNS model from a new angle for strings charged under the Maximal Gauge Symmetry of the  Standard model}
\author{J. S. Bhattacharyya%
\thanks{Electronic address: \texttt{jyoti@kpcoll.ac.in, jtskhrbhattacharyya@gmail.com}}\\
Department of Physics, Kanchrapara College, Kanchrapara 743145,\\  North 24-PGS, West Bengal, India }
\begin{document}
\maketitle
\abstract{We consider the RNS model from a new angle. The longitudinal and time components of the world-sheet fermions add a $U(1)$ charge to a state. Unlike the gauginos, the ground state fermions in the open string sector are complex; spinor representations of $SU(3)_C\otimes SU(2)_L\otimes U(1)_{Y_W}$.} 
\newpage
 In the supersymmetric formulation of $N=1, D=10$ string theory, the GSO conditions are in-built to respect modular invariance and the odd $G$ parity states like tachyons are mapped out. Thus, the ground states in the bosonic sector and the fermionic sector of the open strings are identified as the gauge bosons and the gauginos of a particular chirality. They are connected by a $N=1$ local supersymmetry and both of them transform in the adjoint representation of the gauge group. We argue that the longitudinal and the time components of the world-sheet fermions can add a $U(1)$ charge to the ground state fermions. Since the representations become complex after the addition of the $U(1)$ charge, the gauge group should be unitary and the three generation Standard Model seems to be a good choice.
\par
The light-cone gauge condition: $\psi^+=0$ cannot be taken as an operator identity. It then conflicts with the anti-commutation relation: $\{\psi^+,\psi^-\}=\eta^{+-}$. The physical states can only be annihilated by the positive frequency parts of $\psi^+$ according to Gupta-Bleuler formalism. Let $Y$ be the boson we get after bosonization of the pair of longitudinal and time components of the world-sheet fermions. We interpret $Y$ as the curled up coordinate to define the $U(1)$ gauge symmetry.
\par
Since the spinor representation is complex, the gauge group should be unitary. We assume that the six dimensional internal space is $CP^2\otimes S^2\cong SU(3)_C\otimes SU(2)_L$.
\par
Though $CP^2$ does not admit spinors, $CP^2\otimes S^2$ does \cite{dolan}. It is a well known fact that the index of the Dirac operator on a six dimensional  space of Euclidean signature is $\frac{\chi}{2}$ \cite{witten}, where $\chi$ is the Euler number of the manifold. Since the Euler number of a product manifold is the product of its Euler numbers, the index of the Dirac operator for the product space $CP^2\otimes S^2$ is
\begin{eqnarray}
\nu_+-\nu_-&=&\frac{\chi_{CP^2}.\chi_{S^2}}{2}\\\nonumber
&=& \frac{3.2}{2}\\\nonumber
&=&3
\end{eqnarray}
\footnote{The Betti-Hodge numbers for $CP^n$ are given by
\begin{eqnarray}
b_{pq}&=&\delta_{p,q}\hspace{13pt} p\leq n\\\nonumber
&=&0\hspace{13pt} p> n
\end{eqnarray}
Hence, for $S^2$ or $CP^1$ they are given by the matrix 
\[b_{pq}(S^2)=
\left[ {\begin{array}{cc}
1 & 0 \\
0 & 1 \\
\end{array} } \right]
\]
and those for $ CP^2$ are given by the matrix 
\[b_{pq}(CP^2)=
\left[ {\begin{array}{ccc}
1 & 0 & 0 \\
0 & 1 & 0 \\
0 & 0 & 1 \\
\end{array} }\right].
\]
The same for the product manifold, $M\otimes N$, is given by the Kunneth formula:
\begin{eqnarray}
b_{pq}(M\otimes N)=\sum_{r,m,n,s}b_{rs}(M).b_{mn}(N),
\end{eqnarray}
subject to the constraints $r+m=p$ and $s+n=q$.
Thus, we can write the  Betti-Hodge numbers for $CP^2 \otimes S^2$   as 
\begin{equation}\label{1}
b_{pq}(CP^2\otimes S^2)=
\left[ {\begin{array}{cccc}
1 & 0 & 0 & 0 \\
0 & 2 & 0 & 0 \\
0 & 0 & 2 & 0 \\
0 & 0 & 0 & 1 \\
\end{array} } \right].
\end{equation}
\par
This matrix is homeomorphic to the Hodge matrix of the six dimensional Calabi-Yau Manifold:
\begin{equation}\label{2}
\left[ {\begin{array}{cccc}
1 & 0 & 0 & 1 \\
0 & 3 & 0 & 0 \\
0 & 0 & 3 & 0 \\
1 & 0 & 0 & 1 \\
\end{array} } \right].
\end{equation}
corresponding to three generations of fermions. We have added 1 to each of $b_{11}$ and $b_{22}$ going from \eqref{1} to \eqref{2}. It corresponds to adding one simplex of dimension two and one simplex of dimension four. To keep the Euler characteristics the same, we have added 1 to each of $b_{03}$ and $b_{30}$. It corresponds to the addition of two simplices of dimension three to the manifold.}
\par
Since $\chi$ vanishes if the internal space is a product space of the form $S^1\otimes X$ by the product rule for $\chi s$, there is perhaps no other choice than $CP^2\otimes S^2\neq S^1\otimes X$ for the six dimensional  internal space if the gauge group is unitary.
\par
We can write an arbitrary spinor as
\begin{equation}\label{multiplets}
 \eta=\omega^{(0,0)}\zeta+\omega^{(0,2)}_{\bar{\mu}\bar{\nu}}\gamma^{\bar{\mu}\bar{\nu}}\zeta+\omega^{(0,1)}_{\bar{\mu}}\gamma^{\bar{\mu}}\zeta+\omega^{(0,3)}_{\bar{\mu}\bar{\nu}\bar{\rho}}\gamma^{\bar{\mu}\bar{\nu}\bar{\rho}}\zeta,
\end{equation}
where $\gamma^\mu$-matrices with complex indices $\mu$=1,2 or 3 for the six dimensional internal manifold, can be used as creation and annihilation operators \cite{misra}. The eight components of the spinor break up under $SU(3)$ subgroup of $SU(4)\cong SO(6)$ as $8=4+\bar{4}=1+3+\bar{3}+\bar{1}$. The four states of the four-component  Weyl spinor of $CP^2\otimes S^2$ can be identified with the singlet and the triplet representations (1+3) of the left-handed leptons and quarks of a particular generation. For isoscalar right-handed fermions we should consider the conjugate pair ($\bar{1}+\bar{3}$) of opposite helicity contributing 2.(1+3)=8 degrees of freedom to the fermionic sector. Left-handed fermions are isodoublets. Hence, the corresponding number is sixteen. Therefore, the total number of degrees of freedom of the fermions is twenty-four. The total number of degrees of freedom of the gluons, W bosons, Z bosons and photons is also 2.(8+3+1)=24. So there is an exact matching between the number of degrees of freedom of bosons and fermions at the ground level. This will be so for other states also if the GSO conditions are respected in both the fermionic and the bosonic sectors. Hence, the zero-point energy vanishes and there is no Casimir force between two parallel D-branes \cite{col,pol,pol1}.
\par
The $SU(3)$ quantum numbers are part of the dynamics of the system, but the Chan-Paton factors for the isospin can only be added to the states through the inclusion of Wilson lines and therefore the $SU(2)$ gauge symmetry can break spontaneously. 
\par
We can ignore the group indices locally.  Hence, from the triality symmetry of $SO(8)$, the open string ground states in the fermionic and the bosonic sectors can be identified with copies of the eight primitive states of the Majorana-Weyl fermion and the eight states of the vector meson of the D = 10 N = 1 local supersymmetry.
\par
D-branes cannot be wrapped in the compact internal space in a non-trivial way to add Chan-Paton factors to the states through Wilson loops. Every loop can be shrunk to a point in this case, because $CP^2\otimes S^2\neq S^1\otimes X$. Hence, they should be wrapped in the four dimensional Minkowski space only, making the theory four dimensional in effect. It corresponds to the gauge symmetry 
\begin{eqnarray}
SO(2^\frac{4}{2})&=&SO(4)\\\nonumber
&\cong& SU(2)_L\otimes SU(2)_R.
\end{eqnarray}
The states are $SU(2)_R$ singlets in this case. It confirms that only the $SU(2)$ Chan-Paton factor can be added to the states through D-branes.
\par
Type-I theory has odd-branes, D3 and D1 branes in this case. A D3-brane fills the whole Minkowski space and should correspond to the ordinary Chan-Paton factor of $SU(3)_C$. The Chan-Paton factors of $SU(2)$ introduced on Wilson lines correspond to D1-branes. The magnetic flux enclosed by a Wilson loop through the two dimensional plane normal to the D1-branes is quantized, because $\pi_1(U(1)) =\mathbb{Z}$. The size of the loop is a measure of the energy scale. When  it is small, the size of the loop is sufficiently large to enclose a magnetic flux, breaking the $SU(2)$ gauge symmetry, which is restored when the energy scale is high.
\par
\noindent\rule{\textwidth}{1pt}

\end{document}